\documentstyle[12pt]{article}
\input psfig
\def\be{\begin{equation}}
\def\ee{\end{equation}}
\def\ba{\begin{array}}
\def\ea{\end{array}}
\def\beqn{\begin{eqnarray}}
\def\eeqn{\end{eqnarray}}
\def\bc{\begin{center}}
\def\ec{\end{center}}
\def\bt{\begin{tabular}}
\def\et{\end{tabular}}

\def\vud{$|V_{ud}|$}
\def\vus{$|V_{us}|$}
\def\vcb{$|V_{cb}|$}

\def\vcd{$|V_{cd}|$}
\def\vcs{$|V_{cs}|$}
\def\vtd{$|V_{td}|$}

\def\rub{$|\frac {V_{ub}}{V_{cb}}|$}
\def\mu{$m_u$}

\def\eps{$\varepsilon_K$}
\def\xd{$x_d$}
\def\del{$\delta$}
 
\def\sin2{sin$2\beta$}
\def\b{$\beta$}
\def\etaa{${\bar \eta}$}
\def\rh{${\bar \rho}$}
\def\abg{$\alpha$, $\beta$ and $\gamma$}
\def\dmd{$ \Delta m_d$}
\def\dms{$ \Delta m_s$}
\def\dmds{$ \Delta m_{d,s}$}

\begin{document}
\title{Probing CKM parameters through unitarity, 
$\varepsilon_K$ and $\Delta m_{B_{d,s}}$}
\author{Monika Randhawa and Manmohan Gupta \\
{\it Department of Physics,}\\
{\it Centre of Advanced Study in Physics,}\\
 {\it Panjab University, Chandigarh-
  160 014, India.}}
  \maketitle
\begin{abstract}
A recently carried out unitarity based analysis
\cite{paper2,paper5} involving
the evaluation of Jarlskog's rephasing invariant parameter
J as well as the evaluation of the angles \abg~ of the
unitarity triangle is extended to include the effects of
  $K^{\rm o} - \bar{K^{\rm o}}$ as well as
$B^{\rm o} - \bar{B^{\rm o}}$  mixings. The present analysis
attempts to evaluate CP violating phase \del, \vtd~ and the
angles \abg~ by invoking the constraints imposed by
unitarity,  $K^{\rm o} - \bar{K^{\rm o}}$  and
$B^{\rm o} - \bar{B^{\rm o}}$ mixings independently as well as
collectively.
By invoking the ``full scanning" of input parameters, as used by
Buras {\it et al.}, our analysis yields the following results
respectively for $\delta$ and $|V_{td}|$, 
unitarity: 50$^{\rm o} \pm 20^{\rm o}$ (in I quadrant),
 130$^{\rm o} \pm 20^{\rm o}$ (in II quadrant) 
and  $5.1 \times 10^{-3} \leq |V_{td}|  \leq 13.8 \times 10^{-3}$,
 unitarity and $\varepsilon_k$: 
33$^{\rm o} \leq \delta \leq  70^{\rm o}$ (in I quadrant),
 110$^{\rm o}  \leq \delta \leq   150^{\rm o}$ (in II quadrant), 
$ 5.8 \times 10^{-3} \leq |V_{td}|  \leq 13.6 \times 10^{-3}$,
 unitarity and \dmd:  $30^{\rm o} \leq \delta \leq 70^{\rm o}$, 
 $6.4 \times 10^{-3} \leq |V_{td}|  \leq 8.9 \times 10^{-3}$.
Incorporating all the constraints together, we get
 37$^{\rm o} \leq  \delta \leq 70^{\rm o}$,
$ 6.5 \times 10^{-3} \leq |V_{td}|  \leq  8.9 \times 10^{-3}$ ,
the corresponding ranges  for $\alpha$, $\beta$ and $\gamma$
are, $80^{\rm o} \leq \alpha \leq 124^{\rm 0}$, $15.1^{\rm o} 
\leq \beta \leq 31^{\rm 0}$ and  $37^{\rm o} \leq \gamma \leq 
70^{\rm 0}$, again in agreement with the recent analysis of 
Buras {\it et al.} as well as with 
other recent analyses. Our analysis yields 
0.50 $\leq {\rm sin}2\beta  \leq 0.88$,
in agreement with the the recently updated BABAR results.
This range of sin2$\beta$ is in agreement with the earlier
result of BELLE, 
however has marginal  overlap only with
the lower limit of recent BELLE results.
Interestingly, our  results indicate that the lower limit of
 \del~ is mostly determined
by unitarity, while the upper limit is governed by constraints
 from \dmd.
\end{abstract}
\section{Introduction}
In the context of CKM phenomenology several important developments have
taken place in the last decade. On the one hand, there is considerable
refinement in the data pertaining to CKM parameters, on the other
hand several phenomenological inputs  have also refined the 
CKM elements further. However, despite a good deal of knowledge
 of CKM matrix elements, the CP violation still remains one of the 
least tested aspect of the CKM paradigm. In this context, an intense
amount of experimental activity is being carried out in a variety of 
experiments at B-factories, KEK-B, SLAC-B, HERA-B and will be carried out  
at LHC-B and B-TeV, both of which will become operational by 2006.
 The main goal of the current phase of experiments will be
to observe CP violation in the B system. It is expected
that the data provided by these factories would throw new lights on the
CP violating B decays, in particular it will lead to the construction
of unitarity triangle involving first and third columns of the
CKM matrix.

In this context, the recent flurry of activity created by the
preliminary BABAR \cite{babar} and BELLE \cite{belle} results regarding
the time dependent CP asymmetry $a_{\psi K_S}$
in $B^{\rm o}_d({\bar B}^o_d) \rightarrow \psi K_S$,
has given lot of impetus to the study of CKM phenomenology. Not
withstanding the agreement of the latest BABAR results 
\cite{babarlat}  with the SM (Standard Model)
 \cite{burasrev}-\cite{hocker}, the
activity generated by their preliminary results
has thrown open several interesting possibilities of discovering
new physics beyond the SM. In this context, it has also been
 emphasized by several authors in their recent works
\cite{kagan}-\cite{hurth}  that 
there could be new physics in the loop dominated 
 processes such as $K^{\rm o} - \bar{K^{\rm o}}$ mixing
and $B^{\rm o} - \bar{B^{\rm o}}$ mixing,
 as well as in some of the CP violating amplitudes.

In view of the possibility of signals of new physics in
 $K^{\rm o} - \bar{K^{\rm o}}$ mixing and
  $B^{\rm o} - \bar{B^{\rm o}}$ mixing,
 it is desirable to sharpen their 
implications on the CKM parameters, particularly on the CP violating
phase \del~ as well as on the angles \abg~ 
of the much talked about unitarity triangle.
 In this context,  in the last few years some very detailed
 and extensive analyses of CKM phenomenology
  \cite{gupta}-\cite{peccei}
   have been carried out. These analyses
 primarily  concentrate on finding CKM parameters in the
 Wolfenstein representation \cite{wolf} and the angles 
of unitarity triangle.
 Apart from unitarity, usual inputs for such analyses are CP violating
 parameter $\varepsilon_K$ as well as the
 mass difference, $\Delta m_{d,s}$, between the light
 and heavy mass eigenstates of 
the $B^{\rm o}_{d,s}- \bar{B^{\rm o}_{d,s}}$ system.
 A careful perusal of these analyses indicates that these
do not adequately emphasize the evaluation of Jarlskog's
 rephasing invariant parameter J \cite{jarlskog},
a key quantity measuring the CP violating effects, as well as these
 do not investigate independently the
 implications of  unitarity, $\varepsilon_K$, \dmd~ and
\dms~ respectively.
To closely scrutinize the effects of unitarity,  
$K^{\rm o} - \bar{K^{\rm o}}$  and $B^{\rm o} - \bar{B^{\rm o}}$
 mixing on the CKM parameters,
one should study their implications separately. We 
believe, this would facilitate deciphering the signals of new physics in 
the loop dominated CP violating amplitudes in the K as well as 
B mesons.

Recently Randhawa {\it et al.} \cite{paper2,paper5},
have studied in detail the implications of unitarity on J
and on the CP violating phase \del, as well as
a ``reference triangle'' has  been constructed purely from 
the considerations of unitarity of the CKM matrix.
The purpose of the present communication is to extend the 
analysis carried in references \cite{paper2,paper5} by including
independently the implications of
$K^{\rm o} - \bar{K^{\rm o}}$  and $B^{\rm o} - \bar{B^{\rm o}}$
 mixing, along with the unitarity on 
the CKM parameters. Besides comparing our results with one of the most
 recent analyses \cite{burasrev}, we  also intend to 
  study the effect of the future 
refinements in the data.

The plan of the paper is as follows.         
To facilitate the discussion
as well as to make the mss. readable, in section 2 we 
detail the  essentials of the unitarity based analysis
carried out in reference \cite{paper5}.
In the section 3,
we determine the CKM phase \del~ from unitarity and 
the  $K^{\rm o} - \bar{K^{\rm o}}$ mixing
parameter \eps~ and then use it to evaluate CKM matrix elements 
and  the angles \abg~ of the CKM triangle.
Similarly, in the section 4 we evaluate above mentioned
 CKM parameters using unitarity and the mass differences 
\dmd ~ and \dms.
 In section 5, we discuss our results obtained after incorporating
all the constraints, for example, unitarity, \eps~ and \dmds,
 simultaneously. A comparison of our results
with those of Buras \cite{burasrev} is also included in this section.
 The implications of the 
future refinements in the data
have been  discussed in section 6. 
Finally, section 7 summarizes
our conclusions.
\section{Unitarity and its implications}
Unitarity of the CKM matrix implies nine relations, three 
are diagonal and the six non-diagonal relations can be
 expressed through the six unitarity triangles in the complex 
plane, for example, 
\beqn
  uc& & V_{ud}V_{cd}^{*} + V_{us}V_{cs}^{*} + V_{ub}V_{cb}^{*}~ =~ 0,
\label{uc} \\
db& & V_{ud}V_{ub}^{*} + V_{cd}V_{cb}^{*} + V_{td}V_{tb}^{*}~ =~ 0, 
\label{db} \\
  ds& & V_{ud}V_{us}^{*} + V_{cd}V_{cs}^{*} + V_{td}V_{ts}^{*}~ = ~0, 
\label{ds}  \\
  sb& & V_{us}V_{ub}^{*} + V_{cs}V_{cb}^{*} + V_{ts}V_{tb}^{*}~~ =~ 0,
\label{sb}  \\
  ut& & V_{ud}V_{td}^{*} + V_{us}V_{ts}^{*} + V_{ub}V_{tb}^{*}~ =~ 0,  
\label{ut}  \\
  ct& & V_{cd}V_{td}^{*} + V_{cs}V_{ts}^{*} + V_{cb}V_{tb}^{*}~~ =~ 0,
 \label{ct}
\eeqn
where the letters $`uc'$ etc. represent the corresponding
unitarity triangle for further discussions.
 The areas of all the six triangles (equations \ref{uc}-\ref{ct})
are  equal and the area of any of the unitarity triangle
 is related to Jarlskog's rephasing invariant parameter J as
\be {\rm J} = 2 \times {\rm Area~ of~any~ of~ the~ Unitarity~ Triangle.}
 \label{area} \ee
In principle, J can be evaluated by using above relation through
 any of the six unitarity triangles, however in practice 
 there are  problems associated with
 the evaluation of area of the triangles. Apart from triangle $uc$, 
all  
other triangles involve elements involving $t$ quark, 
which are not experimentally measured as yet.
Apparently, it seems that the triangle $uc$ is ideal for evaluating 
J as it involves the elements known with fair degree of accuracy.
However, on a closer look one finds that 
  $uc$ is a highly squashed triangle, for example,  
the sides $|V_{ud}^*V_{cd}|$
 and $|V_{us}^*V_{cs}|$ are of comparable lengths
 while the third side
  $|V_{ub}^*V_{cb}|$  is several orders of
    magnitude smaller compared to the other two,
hence it is not easy to evaluate its area unambiguously.
Assuming the existence of CP violation through the
CKM mechanism, in reference
\cite{paper2,paper5} we have detailed the procedure for
 evaluation of J through triangle $uc$. Using the experimental
values of CKM matrix elements, listed in the Table \ref{tabinput}, 
we obtain
       \be {\rm J}= (2.59 \pm 0.79) \times 10^{-5}
          \label{jpdg1s}.  \ee

In view of the fact that the triangle is highly squashed, therefore
to ascertain whether J given by equation \ref{jpdg1s}
faithfully incorporates the constraints of unitarity, it
is very much desirable to compare our results in this regard with
those of PDG \cite{pdg}, who have also calculated
CKM matrix by incorporating
unitarity and experimental data pertaining to \vud, \vus, \rub,
\vcd, \vcs~ and \vcb. In the context of PDG CKM matrix, it may
be noted that
once CKM elements are known, one can calculate J corresponding to
any of the six triangles, however the
two unsquashed triangles $db$ and $ut$ provide the best
 opportunity in this regard. The J values 
evaluated from PDG CKM matrix corresponding
 to the triangles $db$ and $ut$ are given as
\beqn {\rm J}_{db} &=& (2.51 \pm 0.87) \times 10^{-5},   \label{jdb} \\
 {\rm J}_{ut} &=& (2.45 \pm 0.91) \times 10^{-5}.    \label{jut}  \eeqn
 The agreement between the equations \ref{jpdg1s}, \ref{jdb} and
 \ref{jut} strongly supports our evaluation of J through the
  triangle $uc$
 as detailed in \cite{paper2,paper5}.

To evaluate \del~ from J, we first consider the 
 standard  representation of CKM matrix \cite{pdg}, for example,
 \be V_{{\rm CKM}}= \left( \ba {lll} c_{12} c_{13} & s_{12} c_{13} &
  s_{13}e^{-i\delta} \\
  -s_{12} c_{23} - c_{12} s_{23} s_{13}e^{i\delta} &
 c_{12} c_{23} - s_{12} s_{23}s_{13}e^{i\delta}
  & s_{23} c_{13} \\
  s_{12} s_{23} - c_{12} c_{23} s_{13}e^{i\delta} &
  - c_{12} s_{23} - s_{12}c_{23} s_{13}e^{i\delta} &
  c_{23} c_{13} \ea \right),  \label{ckm}  \ee
  with $c_{ij}={\rm cos}\theta_{ij}$ and
   $s_{ij}={\rm sin}\theta_{ij}$ for
 $i,j=1,2,3$. In terms of the CKM elements, 
J can be expressed as \cite{jarlskog} 
\be {\rm J} = s_{12}s_{23}s_{13}c_{12}
        c_{23}c^2_{13}{\rm sin} \delta. \label{j} \ee
Calculating $s_{12}$, $s_{23}$ and $s_{13}$ from \vus, \vcb~
and \rub,  listed in the table \ref{tabinput}, one
can plot a distribution for \del, using 
 equations \ref{jpdg1s} and \ref{j}  yielding
 \be \delta = 50^{\rm o} \pm 20^{\rm o}, \label{del1} \ee
which in the second quadrant translates to
\be \delta = 130^{\rm o} \pm 20^{\rm o}. \label{del2} \ee
This value of $\delta$ apparently looks to be the consequence
  only of the unitarity relationship given by equation \ref{uc}.
    However, on further investigations, as shown by Branco and Lavoura
  \cite{branco}, one finds that this $\delta$ range is a consequence
   of all the non trivial unitarity constraints. In this sense, the above
  range could be attributed to the unitarity of the
 CKM matrix. 

Having found $\delta$, it is desirable to examine whether its above range
 and the mixing angles
$s_{12}$, $s_{23}$ and $s_{13}$,
  as calculated from the experimental
values of \vus, \vcb~ and \rub, are able to  reproduce the CKM matrix 
given by PDG. The  CKM matrix thus evaluated is given as
  \be   \left[ \ba {ccc}   0.9751-0.9761  &  0.2173-0.2219  &
  0.0025-0.0048 \\
0.2169-0.2219 & 0.9742-0.9754 & 0.0383-0.0421 \\
  0.005-0.0138  &  0.0365-0.0420 &   0.9991-0.9993  \ea \right].
    \label{unickm} \ee
The above matrix has a good deal of overlap with that of PDG matrix.
  To facilitate a more detailed  comparison, we present below the
  PDG matrix \cite{pdg}, 
   \be   \left[ \ba {ccc}  0.9742-0.9757 & 0.219-0.226 & 0.002-0.005 \\
    0.219-0.225 & 0.9734-0.9749 &  0.037-0.043 \\
    0.004-0.014 &  0.035-0.043 & 0.9990-0.9993 \ea \right].
     \label{ckmpdg} \ee
      Comparison of \ref{unickm} and \ref{ckmpdg} reveals that for 
most of the elements, we have an excellent overlap,
   justifying the procedure followed in carrying out the
  present analysis. It should be noted that there
  is an excellent agreement in the case of \vtd,
 the element having a sensitive dependence on $\delta$.

Having found $\delta$, one can evaluate the angles 
 $\alpha$, \b~ and $\gamma$ of the unitarity triangle $db$,
 likely to be measured in the near future.
The angles of the  triangle can be
expressed  in terms of the CKM elements as
\beqn \alpha& =& {\rm arg}\left(\frac{-V_{td}V_{tb}^*}{V_{ud}V_{ub}^*} 
\right), 
\label{alpha} \\
 \beta& =& {\rm arg}\left(\frac{-V_{cd}V_{cb}^*}{V_{td}V_{tb}^*}\right),
 \label{beta} \\
\gamma &=& {\rm arg}\left(\frac{-V_{ud}V_{ub}^*}{V_{cd}V_{cb}^*}\right).
 \label{gamma} \eeqn
 Making use of the PDG representation of CKM matrix,
 experimental values of  \vus, \vcb~ and \rub~ 
 from table \ref{tabinput} and the range of
\del, given by equations \ref{del1} and \ref{del2},
 one can easily find out 
the corresponding ranges for the three angles, listed
in column II of Table \ref{tab1}.

 The range of  \sin2, corresponding to \del~ given by
equations \ref{del1} and \ref{del2}, is
 \be {\rm sin}2\beta =  0.28~ {\rm to} ~0.88\,, \label{sin2uni} \ee
which is in good agreement with the
latest  BABAR results \cite{babarlat}, for example, 
 \be {\rm sin}2\beta = 0.59 \pm 0.14 \pm 0.05 \qquad {\rm BABAR}.
  \label{babarlat} \ee
This range of sin2$\beta$ is in agreement with the earlier
result of BELLE, 
however has marginal  overlap only with
the lower limit of recent BELLE results 
\cite{bellelat}, for example,
\be {\rm sin}2\beta = 0.99 \pm 0.14\pm 0.06 \qquad {\rm BELLE}.
  \label{bellelat} \ee

Most of the analyses of the unitarity triangle have been
carried out using the Wolfenstein parametrization \cite{wolf},
therefore to facilitate the comparison of our results with
other contemporary analyses, we have also evaluated \rh~ and \etaa~
defined as
\be {\bar \rho} = \rho\left(1-\frac{\lambda^2}{2}\right),~~~~~~
  {\bar \eta} = \eta\left(1-\frac{\lambda^2}{2}\right)
 \label{rhoeta} \ee
where $\rho = \frac{s_{13}}{s_{12}s_{23}}{\rm cos}~\delta, 
 ~\eta = \frac{s_{13}}{s_{12}s_{23}}{\rm sin}~\delta$ and
 $\lambda \simeq s_{12}$. It may be mentioned that while
evaluating  $\alpha$, \b, $\gamma$, \rh~ and \etaa, we have used
the method of ``full scanning'' of the input parameters
as employed by Buras {\it et al.} \cite{burasrev}.
Our results in this regard are listed in the Table \ref{tab1}.
\section{$K^{\rm o} - \bar{K^{\rm o}}$ mixing and its implications}
From the previous section, it is clear that unitarity
alone doesn't give any significant constraints on \del~ and \vtd.
However, in  view of the fact that 
the CP violating parameter \eps~  
is experimentally very well determined, it is therefore
interesting to examine its implications on \del~ and \vtd.
To this end, we consider the following QCD corrected and quark loop
dominated expression for \eps~ \cite{buchalla},
\be |\varepsilon_K| = \frac{G_F^2 F_K^2 m_K m_W^2}
        {6 \sqrt{2} \pi^2 \Delta m_k}B_K{\rm Im}\lambda_t
  [{\rm Re} \lambda_c(\eta_1 F(x_c)-\eta_3 F(x_c,x_t)) 
 - {\rm Re} \lambda_t \eta_2 F(x_t)], \label{eps} \ee
where $F(x_i)$ are Inami-Lim functions, $x_i=m_i^2/M_W^2$ and
$\lambda_i=V_{id}V^*_{is}$, $i=c,t$.
In terms of the mixing angles and the phase \del,
Im$\lambda_t$,
Re$\lambda_t$ and Re$\lambda_c$ can be expressed as
\beqn {\rm Im}\lambda_t & = &
 s_{23}s_{13}c_{23}{\rm sin}\delta,  \\
{\rm Re}\lambda_t & = &
 s_{23}s_{13}c_{23}(c_{12}^2-s_{12}^2){\rm cos}\delta - 
s_{12}c_{12}(s_{23}^2-c_{23}^2s_{13}^2),\\
{\rm Re}\lambda_c & = &
 s_{23}s_{13}c_{23}(s_{12}^2-c_{12}^2){\rm cos}\delta - 
s_{12}c_{12}(c_{23}^2-s_{23}^2s_{13}^2). \eeqn
To examine the exclusive role played by \eps~ 
in restricting \del, we have calculated \eps~ by 
varying \del~ from 0$^o$ to 180$^o$ and by `` full
scanning'' of the various
 input parameters.
 The experimental limits on \eps, for example
\be \varepsilon_K = (2.280 \pm 0.013) \times 10^{-3},
 \label{epsexpt} \ee
yield \del~ in the range
\be \delta = 33^o~ {\rm to}~ 163^o. \label{delonlyeps} \ee
The above range of \del~ when combined with the constraints
obtained in the last section from unitarity becomes
\beqn \delta &=& 33^{\rm o} ~ {\rm to}~ 70^{\rm o}
~~~~({\rm in~ I~ quadrant}), \label{deleps1} \\
 & & 110^{\rm o} ~{\rm to} ~150^{\rm o}
 ~~~~({\rm in~ II~ quadrant}). \label{deleps2} \eeqn
In the Table \ref{tab1}, we have listed     
the ranges for \vtd, $\alpha$, $\beta$, $\gamma$,  \rh~ and \etaa~
obtained from the unitarity constraints,
the results corresponding to the combined constraints of unitarity
and \eps~ are presented in column III of the same table.

There are several important points which need to be discussed
further. It is interesting to note that even a well determined \eps,
despite its direct proportionality to sin$\delta$,
is not able to improve upon
the range of \del~ implied by unitarity.
This could be understood presumably by closely examining
 the expression for \eps, which involves parameters such as 
$B_K$, \rub~ and \vcb,
having much larger uncertainties than that of \eps.

\begin{figure}[hbt]
\centerline{\psfig{figure=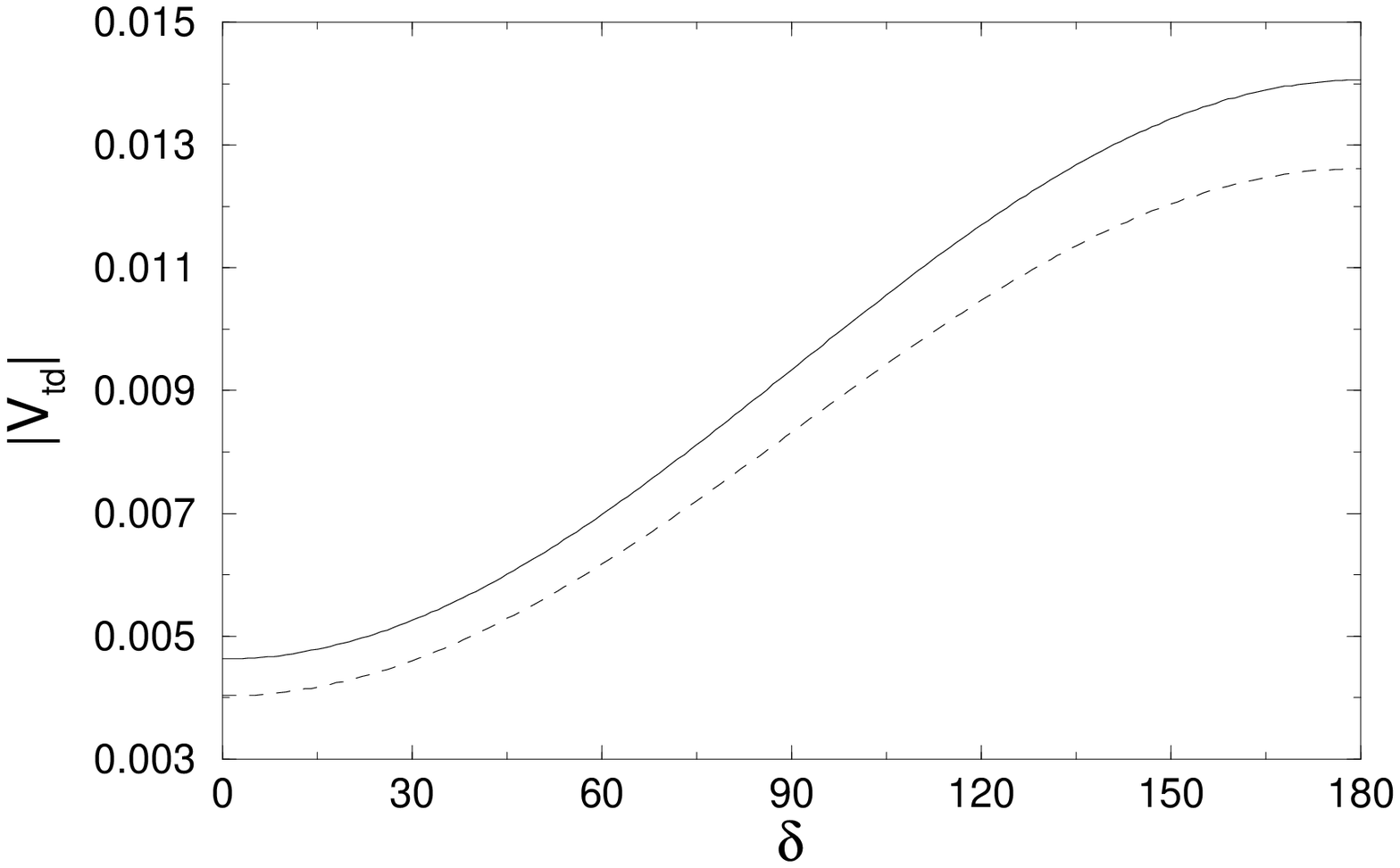,width=4in,height=3in}}
   \caption{Plot of \del~ versus $V_{td}$. The solid line
   corresponds to maximum value of \vtd~ for a given \del~
   and the broken line corresponds to
    minimum value of \vtd~ for a given \del.}
\label{figvtd}   
\end{figure}

Similarly, one finds that \eps~ is not able
to constrain \vtd~ much in comparison to the unitarity.
This  looks to be a consequence of not so improved constraints
 on \del~ in comparison with the unitarity.
 To examine this more closely, we consider the expression
 for  $V_{td}$
 in the standard parametrization, for example,
\be V_{td}=s_{12} s_{23} - c_{12} c_{23} s_{13}e^{i\delta}.
\label{vtd}   \ee
 In figure \ref{figvtd}, we have plotted
 \del~ versus \vtd, depending upon $s_{12},~s_{23},~ s_{13}$
and $\delta$,
 the solid curve corresponding to maximum \vtd~
 and the other one corresponding to minimum \vtd.
It is clear from the figure
 that for low values of \del~ in the first quadrant and 
correspondingly for large values of \del~ in the second 
quadrant, \vtd~ is not very sensitive to variations in \del.
For example, examining the solid curve, one finds that
variation in \del~  from $0^{\rm o}$
to 20$^{\rm o}$ corresponds to \vtd~ from
 0.0046 to 0.0049, while in the case when \del~ varies from
 70$^{\rm o}$ to 90$^{\rm o}$, there is a sharp variation in 
 \vtd, for example it varies from 0.0078 to 0.0094.
 Similar conclusions follow from the other curve.
  Hence \eps~ constraint is not able to register any noticeable
improvement in the range of \vtd~ in comparison to the unitarity constraints.

\begin{figure}
\centerline{\psfig{figure=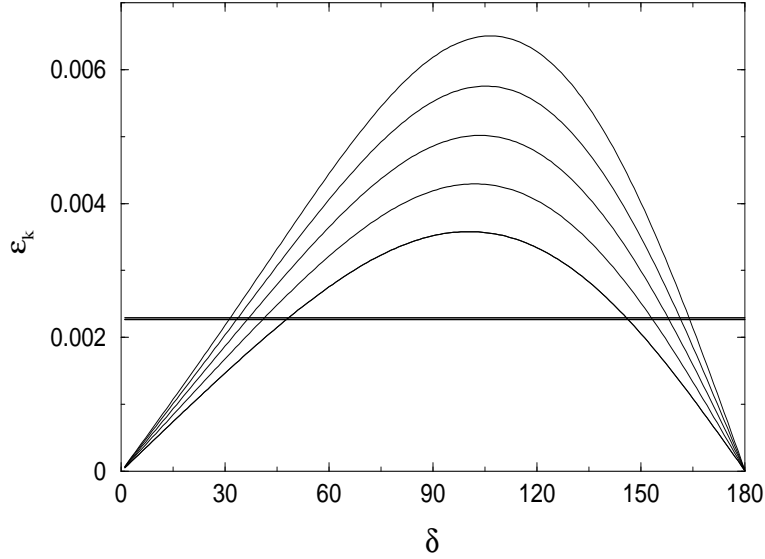,width=4in,height=3in}}
   \caption{Plot of  $\delta$ versus \eps~ for \rub = 0.065, 0.078,
 ........., 0.115, keeping other parameters fixed.
The horizontal lines represent the experimental limits
of \eps~ given by equation \ref{epsexpt}.}
\label{figvcb}  
 \end{figure}

After having examined the reasons for not so sharp limits on \del~
and \vtd~ by \eps, it may be of interest to investigate which
parameters are primarily responsible for this.
We have already noted that 
\eps~ depends on $B_K$, \rub~ and \vcb~ etc.,
it is therefore likely that the  uncertainties in $B_K$ and the CKM elements 
\rub~ and \vcb~ might be playing an important role in giving
not so sharp limits on \del.
To examine this point further, in figure \ref{figvcb} we
have plotted \eps~ versus \del~ scanning full range of \rub,
keeping other parameters  fixed.
The uppermost curve is for \rub=0.115
and the lowermost curve corresponds to \rub=0.065, the curves
corresponding to other values of 
 \rub~  lying in between.
 Horizontal lines represent the experimental range for \eps.
 From the figure, it is evident that there is no value of \rub~
 which is excluded by \eps~ constraints. Further, it is
 also clear from the figure that the refinements in
\rub~ would not lead to any significant improvements in
 the range of \del~ beyond that given in equation
\ref{delonlyeps}, however if the upper limit of \rub~
comes down, then the lower limit of \del~ will go up slightly. 
In the same manner, one can find that for the present experimental range
of \vcb, no value of \vcb~ is
excluded by the \eps~ constraints, as well as further refinements
in \vcb~ will not affect the \del~ range significantly. 
From the above discussion, 
one may conclude that the reason for not so sharp cuts on \del~
basically emanates from the `large uncertainties' in $B_K$.
\section{$B^{\rm o} - \bar{B^{\rm o}}$ mixing and its implications}
We have already seen that \eps~ and unitarity are not able to
constrain \del~ much. However, it is well known that the  
$B^{\rm o} - \bar{B^{\rm o}}$ mixing is a loop dominated process with 
major contributions coming from the $t$ quark, consequently
from $V_{td}$. Therefore
$B^{\rm o}_d - \bar{B^{\rm o}_d}$ mixing parameter \xd, related to
 the mass difference $\Delta m_d$, is expected to yield significant
 constraints on \vtd.
 To this end, we consider
the QCD corrected expression for $\Delta m_d$ \cite{buchalla},
for example
\be \Delta m_d = \frac{G_F^2}{6 \pi^2} \eta_{\tiny{QCD}}
m_{B_d}B_{B_d} F^2_{B_d} M^2_W F(x_t)|V_{td}|^2.
 \label{xd} \ee 
Using the input values of various parameters listed in table
 \ref{tabinput}, and carrying
out full scanning of all the parameters, we obtain 
\be |V_{td}| = 0.0064~ {\rm to} ~0.0089. \label{vtdxd} \ee
Comparing this range with that obtained from unitarity
and \eps~ given in table \ref{tab1},
we find that \dmd~ considerably improves the constraints on \vtd.
The above improved constraints on \vtd~ can be used to yield
 constraints on \del.
Using \vus, \vcb~ and \rub, listed in the table \ref{tabinput},
 along with the  equation \ref{vtd} and above range of
\vtd, we obtain the following limits on \del
\be \delta = 1^{\rm o}~ {\rm to}~  104^{\rm o}.
 \label{delonlyxd} \ee
It may be noted that the above range of \del~ is 
exclusively a consequence of \dmd~ constraint.
This range when combined with the unitarity constraints, obtained
in section 2, yields
\be \delta = 30^{\rm o}~ {\rm to}~  70^{\rm o}.
 \label{delxd} \ee 
Thus, the combined constraints from \dmd~ and
 unitarity limit \del~ to the first quadrant.
The corresponding ranges of $\alpha$, $\beta$, $\gamma$,
\rh~ and \etaa~ are listed in the column IV of Table \ref{tab1}.

For the sake of completeness, we have also investigated whether
 \dms~ could also constrain \del. In this regard,
  we find that the presently available lower
limit on \dms~ (\dms $>$ 15.0 $ps^{-1}$ ) 
does not lead to any appreciable 
constraints on \del~ and other quantities mentioned above.
This is because of the fact that in \dms, the \del~
 dependence comes through $V_{ts}$, an element which is not
very sensitive to variations in \del.
 However, our calculations reveal that the lower bound on
 \dms~ yields a constraint on the upper bound of \del,
 for example, if in future \dms~ is found to be greater than
 27ps$^{-1}$, then  one can show that this would imply
\del$\leq 138^o$. Therefore, until and unless one finds a
remarkable change in the lower limit of \dms, one may
not be able to improve upon a much better bound on \del~ due to \dmd.
\section{Unitarity, \eps~ and $\Delta m_{d,s}$ together}
In most of the past analyses of the CKM phenomenology the 
constraints from unitarity, \eps, \dmd~ and \dms~  have been
simultaneously applied to obtain \rh, \etaa~ and the CP angles
 $\alpha$, $\beta$ and $\gamma$. To facilitate comparison
 of our results with other recent analyses
  \cite{burasrev,parodi,hocker} 
 as well as to construct the unitarity
triangle by incorporating all the available constraints, we have
also carried out an analysis wherein the implications of unitarity,
$K^{\rm o} - \bar{K^{\rm o}}$ mixing and $B^{\rm o} - \bar{B^{\rm o}}$ mixing have been
studied together. The corresponding results have been listed in
the last column of the Table \ref{tab1}. A closer look at the
Table \ref{tab1} reveals several interesting points. The combined
 constraints of unitarity, $K^{\rm o} - \bar{K^{\rm o}}$ mixing and 
$B^{\rm o} - \bar{B^{\rm o}}$ mixing force \del~ to lie in the range 
$37^{\rm o}~ {\rm to}~  70^{\rm o}$, implying
a very restricted range for \vtd, for example
 \be |V_{td}|= 0.0065~ {\rm to} ~0.0089. \label{vtdall} \ee
The corresponding range of \sin2~ is
 \be {\rm sin2}\beta =  0.50~ {\rm to} ~0.88, \label{sin2final} \ee
which is almost within the latest BABAR results, however
the lower limit is slightly above that of BABAR.
In case the recent BELLE result (equation (\ref{bellelat})) 
is confirmed, it will have serious implications for CKM paradigm
as well as the origin of CP violation in the Standard Model. 
These issues will be discussed in detail elsewhere. 
It may be of interest to mention that when we carry a full scanning
of $s_{12}$, $s_{23}$
and $s_{13}$, we find that ${\rm sin}2\beta~\leq~0.88$ for
 all values of \del, the equality is achieved for
  \del$\sim 60^{\rm o}$.
Therefore, in case the upper limit of \sin2~ is found to be
less than 0.88, it would imply \del~ being less than $60^{\rm o}$
having implications for \vtd.

Coming to the question of comparison of our results with
those of other recent analyses, our results can best be
compared with those of Buras {\it et al.} \cite{burasrev},
as we have also used the ``full scanning" approach for
inputs.
In the table \ref{tab1}, we have listed their results as well as ours.
On comparing, one finds that the results obtained by us
are almost in agreement with their results. In fact,
except \rh, all other parameters have excellent overlap with
their results. This lends strong support to our evaluation of J and
\del, which has not been carried out earlier.

A critical perusal of our calculations reveals several interesting points.
From the table \ref{tab1},
  one finds that the lower limit of \del~ is mostly determined
by unitarity, constraints from \eps~ push the lower limit on \del~ 
slightly up, while constraints from \dmd~ are instrumental in determining 
the upper limit of \del. In particular, considering
 unitarity and \dmd simultaneoulsy,
 \del~ is restricted to the first quadrant only.
\section{Implications of  refinements in data}
In the last section, we have seen that the combined constraints
of unitarity, $K^{\rm o} - \bar{K^{\rm o}}$ mixing and
 $B^{\rm o} - \bar{B^{\rm o}}$ mixing
have significant impact in constraining \del~ and consequently other 
CKM elements, particularly  \vtd.
It, therefore, becomes interesting to examine the implications 
of future refinements in the data on \del~ and other quantities
mentioned in table \ref{tab1}.

The unitarity analysis involves the  
 elements of first two rows of CKM matrix, all of which are
 known with the reasonable accuracy except for \rub~ and \vcs, where 
the uncertainties are more than 15$\%$. Therefore, it would be 
interesting to examine the consequences of the future refinements in these
 two elements. In this context, we have repeated
the unitarity based analysis with the future value of
\rub, \rub=0.090 $\pm$ 0.010, whereas for \vcs, we consider the 
latest LEP value $|V_{cs}|= 0.996\pm 0.013$ \cite{lep1}. 
The corresponding results for \del, \vtd, $\alpha$, \b, $\gamma$, \rh~
and  \etaa~ are summarized in the Table \ref{tabfuture}.
From the table, one finds that the future refinements in \rub~ and
\vcs~ may result in the relatively stronger bounds on the quantities
mentioned above. In particular, the lower limit of \vtd~ and 
\sin2~ goes up noticeably.
 
As has already been mentioned, the \eps~ analysis with
 the future refinements of data will not have
 significant implications. However, in the case
of \dmd~ a refinement in $F_{B_d}\sqrt{B_{B_d}}$ to the
value  0.230 $\pm$ 0.010 GeV, along with
refinement in  \rub~ mentioned above, 
 has quite marked implications on
\vtd~ and $\delta$.
In particular, $\delta$ range will improve from 
 $1^{\rm o} - 104^{\rm o}$
to  $38^{\rm o} - 80^{\rm o}$, the corresponding improved
range for \vtd~ is, 0.0070 to 0.0084.
In the table \ref{tabfuture}, we have listed the ranges for 
\del~, \vtd,  $\alpha$, $\beta$, $\gamma$, \rh~ and
\etaa~ found by using future refinements of data
corresponding to different combinations of constraints due to
unitarity, \eps~  and \dmds.

It is of interest to discuss the implications of the refinements in the
 measurement of \sin2.
In this context, it may be noted that the lower and
upper limit on \del,
 given by  equation \ref{delxd}, can be attributed respectively to  
unitarity and \dmd.
One can easily check that with \del~ in the first quadrant,
the lower and upper limits of \sin2~ correspond to the lower and upper
limits of \del~ respectively.
Therefore, refinements in the upper limit of \sin2~
will have significant implications for \dmd, whereas the
 lower limit will have implications for unitarity.
\section{Summary and conclusions}
A recently carried out unitarity based analysis
\cite{paper2,paper5} involving
the evaluation of Jarlskog's rephasing invariant parameter
J as well as the CP violating phase \del~
is extended to include the effects of
  $K^{\rm o} - \bar{K^{\rm o}}$ and
$B^{\rm o} - \bar{B^{\rm o}}$ mixings. The present analysis
attempts to evaluate CP violating phase \del, \vtd~ and the
angles of the unitarity triangle 
\abg~ by invoking the constraints imposed by
unitarity,  $K^{\rm o} - \bar{K^{\rm o}}$  and
$B^{\rm o} - \bar{B^{\rm o}}$ mixings independently as well as
collectively.
As discussed in references \cite{paper2,paper5}, the unitarity
constraints have been included via J evaluated through 
the triangle $uc$.
Using the \del~ range, found from the constraints of unitarity
only, we are able to reproduce the PDG CKM matrix
rather well, also found from
the considerations of unitarity.
It may be mentioned that we have used the method of ``full scanning''
 of the input parameters, also  employed by  Buras {\it et al.}
 \cite{burasrev}. The
$K^{\rm o} - \bar{K^{\rm o}}$ mixing and
 $B^{\rm o} - \bar{B^{\rm o}}$ mixing constraints
  have been incorporated through
$\varepsilon_K$ and \dmds~ respectively. The implications of above
mentioned phenomena have been studied one by one 
on the CKM parameters, in particular on the CP violating phase \del~
and \vtd.  This  enables one to  explore 
their  role clearly and independently of each other in constraining 
  $\delta$ and consequently \vtd~ and the angles
 $\alpha$, $\beta$ and $\gamma$ of the unitarity triangle $db$.
Finally we incorporate all the constraints together in order
to compare our result with the other recent analyses, in particular
with Buras {\it et al.}.

Using the latest data listed in table \ref{tabinput}
and incorporating the unitarity constraints, 
we get  $\delta$ = 
50$^{\rm o} \pm 20^{\rm o}$ (in I quadrant),
 130$^{\rm o} \pm 20^{\rm o}$ (in II quadrant) 
and   $5.1 \times 10^{-3} \leq |V_{td}| 
 \leq 13.8 \times 10^{-3}$.
The combined constraints of unitarity and $\varepsilon_k$
yield, 
33$^{\rm o} \leq \delta \leq 70^{\rm o}$ (in I quadrant),
 110$^{\rm o} \leq \delta \leq  150^{\rm o}$ (in II quadrant) and 
$ 5.8 \times 10^{-3} \leq |V_{td}|  \leq 13.6 \times 10^{-3}$,
 thus slightly improving  only the lower limit of \del~ 
over and above unitarity. However, the combined constraints of  
 unitarity and \dmd~ restrict \del~ and \vtd~ significantly,
for example,
$30^{\rm o} \leq \delta \leq 70^{\rm o}$ and 
 $6.4 \times 10^{-3} \leq |V_{td}|  \leq 8.9 \times 10^{-3}$.
  We find that the presently available lower
limit on \dms~ (\dms $>$ 15.0 $ps^{-1}$) 
doesn't yield any significant constraints on $\delta$.  
Finally, incorporating all the constraints together we get,
 37$^{\rm o} \leq \delta \leq 70^{\rm o}$,
$ 6.5 \times 10^{-3} \leq |V_{td}|  \leq  8.9 \times 10^{-3}$ ,
and the corresponding ranges  for
$\alpha$, $\beta$ and $\gamma$
are, $80^{\rm o} \leq \alpha \leq 124^{\rm o}$, $15.1^{\rm o} 
\leq \beta \leq 31^{\rm o}$ and  $37^{\rm o} \leq \gamma \leq 
70^{\rm o}$, which are in agreement with the recent results of
Buras {\it et al.}. 
Also, the calculated range of sin2$\beta$ is
 within the recently updated BABAR results.
This range is in agreement with the earlier
result of BELLE, 
however has marginal  overlap only with
the lower limit of recent BELLE results.

We have also examined the implications of future refinements in data
and have found that the future refinements in \rub, \vcb, \vcs~
and $F_{B_d}\sqrt{B_{B_d}}$ etc.
would considerably narrow the range of \del~  and consequently of \vtd.
Since, with \del~ in the first quadrant,
the lower and upper limits of \sin2~ correspond to the lower and upper
limits of \del~ respectively, therefore, the refinements in
 the measurements of upper limit and the lower limit of \sin2~
will have implications for \dmd~
and unitarity of the CKM matrix respectively.

\vskip 1cm
  {\bf ACKNOWLEDGMENTS}\\
  M.G. would like to thank S.D. Sharma for useful discussions.
 M.R. would like to thank CSIR, Govt. of India, for
 financial support and also the Chairman, Department of Physics,
for providing facilities to work in the department.

\begin{table}
\begin{tabular}{|l|l|c|} \hline
Parameter & Value & Reference \\ \hline
$\frac{G_F^2 F_K^2 m_K m_W^2}
   {6 \sqrt{2} \pi^2 \Delta m_k}$ & 3.84 $\times 10^4$
   & \cite{burasrev} \\
$ F_{B_d}\sqrt{B_{B_d}}$ & $230 \pm 25 \pm 20$ MeV  & \cite{parodi} \\
 \vud & 0.9735 $\pm$ 0.0008 & \cite{pdg} \\
 \vus &  0.2196 $\pm$ 0.0023 & \cite{pdg} \\
\vcd & 0.224 $\pm$ 0.016  & \cite{pdg} \\ 
\vcs & 1.04 $\pm$ 0.16  & \cite{pdg} \\
 \vcb & 0.0402 $\pm$0.0019  & \cite{pdg} \\
 \rub &  0.090 $\pm$ 0.025  & \cite{pdg} \\
$m_W$ & 80.419 $\pm$ 0.056 GeV  & \cite{pdg} \\ 
  $m_t$ & 167 $\pm$ 5 GeV  & \cite{pdg} \\
$ m_c$ & 1.3 $\pm$ 0.1 GeV  & \cite{pdg} \\ 
$ m_{B_d}$ & 5279.2 $\pm$ 1.8 MeV  & \cite{pdg} \\
$ B_K$  & 0.87 $\pm$ 0.06 $\pm$ 0.13  & \cite{lel}  \\ 
$\eta_1$ & 1.38 $\pm$ 0.53  & \cite{hn} \\
$\eta_2$ &  0.574 $\pm$ 0.004  & \cite{hn} \\ 
$\eta_3$ & 0.47 $\pm$ 0.04  &  \cite{hn} \\
$\eta_{QCD}$ & 0.55 $\pm$ 0.01  &  \cite{hn} \\ 
 $\Delta m_d$ & 0.487 $\pm$ 0.014 $\pm$ 0.032  ps$^{-1}$
 & \cite{lep} \\
 $\Delta m_s$ & $>$ 15.0 ps$^{-1}$ at 95$\%$ C.L.  
 & \cite{lep} \\
\hline
\end{tabular}
\caption{Values of the input parameters used for the analysis.}
\label{tabinput}
\end{table}

\begin{table}
\begin{tabular}{|l|l|l|l|l|c|} \hline
 & \bt{c} Using only \\ unitarity \\ \et & 
 \bt{c} Unitarity \\ and  $\varepsilon_K$ \\ \et & 
 \bt{c} Unitarity  \\  and $\Delta m_d$ \\ \et
& \bt{c} Unitarity, \\$\varepsilon_K$ and $\Delta m_d$ \\ \et 
& \bt{c} Buras's\\ results \\ \et \\  \hline
  $\delta$ & \bt {c} 50$^{\rm o} \pm 20^{\rm o}$,\\
 130$^{\rm o} \pm 20^{\rm o}$ \et &
\bt{c} 33$^{\rm o}$ to 70$^{\rm o}$,\\
  110$^{\rm o}$ to 150$^{\rm o}$ \\ \et &
 $30^{\rm o}$ to $70^{\rm o}$  &
 $37^{\rm o}$ to $ 70^{\rm o}$ & -  \\&  &  & && \\
  $|V_{td}|$ & $ \ba{c}(5.1~ {\rm to}~ 13.8) \\ \times~ 10^{-3}\ea $  & 
$ \ba{c}(5.8 ~{\rm to}~ 13.6) \\ \times~10^{-3} \ea$  &
 $\ba{c} (6.4~{\rm to}~ 8.9) \\ \times~10^{-3} \ea $
& $\ba{c} (6.5~ {\rm to}~ 8.9) \\ \times~10^{-3} \ea$
& $\ba{c} (6.7~ {\rm to}~ 9.3) \\ \times~10^{-3} \ea$ \\&& & &  & \\
   $\alpha$ & $20^{\rm o}$ to $139^{\rm o}$ &
$ 20^{\rm o}$ to $124^{\rm o}$ & $80^{\rm o}$ to $139^{\rm o}$ 
&  $80^{\rm o}$ to $124^{\rm o}$
&  $78.8^{\rm o}$ to $120^{\rm o}$\\&& & &  & \\
   $\beta$ & $6.5^{\rm o}$ to $31^{\rm o}$  & 
 $7.4^{\rm o}$ to $31^{\rm o}$ & $ 11^{\rm o}$ to $31^{\rm o}$
& $15^{\rm o}$ to $31^{\rm o}$ 
&  $15.1^{\rm o}$ to $28.6^{\rm o}$ \\ & &&  & & \\
  $\gamma$ & \bt {c} 50$^{\rm o} \pm 20^{\rm o}$,\\
 130$^{\rm o} \pm 20^{\rm o}$ \et &
\bt{c} 33$^{\rm o}$ to 70$^{\rm o}$,\\
  110$^{\rm o}$ to 150$^{\rm o}$ \\ \et &
 $30^{\rm o}$ to $70^{\rm o}$  &
 $37^{\rm o}$ to $ 70^{\rm o}$ &
  $37.9^{\rm o}$ to $ 76.5^{\rm o}$  \\&  &  & && \\
  ${\bar \rho}$ &- 0.447 to 0.447 & - 0.447 to 0.433 &
  0.097 to 0.387  &   0.097 to 0.387 
& 0.06 - 0.34 \\&& & &  & \\
  ${\bar \eta}$ & 0.143 to 0.486 & 0.159 to 0.486 &
0.143 to 0.485  & 0.225 to 0.486
& 0.22 - 0.46 \\
\hline
\end{tabular} 
\caption{Present results regarding $\delta$, $|V_{td}|$, $\alpha$,
 $\beta$, $\gamma$,  ${\bar \rho}$
and ${\bar \eta}$ obtained by using the inputs given in Table 
\ref{tabinput} for different combinations of constraints.}
\label{tab1}
\end{table}

\begin{table}
\begin{tabular}{|l|l|l|l|l|} \hline
 & \bt{c} Using only \\ unitarity \\ \et & 
 \bt{c} Unitarity \\ and  $\varepsilon_K$ \\ \et & 
 \bt{c} Unitarity  \\ and $\Delta m_d$ \\ \et
& \bt{c} Unitarity, \\$\varepsilon_K$ and $\Delta m_d$ \\ \et 
\\ \hline
  $\delta$ & \bt {c} 60$^{\rm o} \pm 18^{\rm o}$,\\
 120$^{\rm o} \pm 18^{\rm o}$ \\ \et &
\bt {c} 42$^{\rm o}$ to 78$^{\rm o}$, \\
 102$^{\rm o}$ to 138$^{\rm o}$ \\ \et &
$42^{\rm o}$ to $ 78^{\rm o}$  & $42^{\rm o}$ to $ 78^{\rm o}$ 
 \\  &&  & & \\
  $|V_{td}|$ & $\ba{c}(6.4 ~{\rm to}~ 12.9) \\ \times~10^{-3} \ea$ & 
$ \ba{c} (6.6~ {\rm to}~ 12.1) \\ \times~10^{-3} \ea$ & 
$ \ba{c} (7.0 ~{\rm to}~  8.4) \\ \times~10^{-3} \ea$ &
$ \ba{c} (7.0~{\rm to}~  8.4) \\ \times~10^{-3} \ea $ \\&& & &   \\
   $\alpha$ & $29^{\rm o}$ to $128^{\rm o}$ &
$ 29^{\rm o}$ to $118^{\rm o}$ & $76^{\rm o}$ to $120^{\rm o}$ 
&  $76^{\rm o}$ to $117^{\rm o}$\\&& &   & \\
     $\beta$ & $10^{\rm o}$ to $27^{\rm o}$  & 
$11^{\rm o}$ to $27^{\rm o}$ & $ 18^{\rm o}$ to $27^{\rm o}$
& $18^{\rm o}$ to $27^{\rm o}$ \\& &&   & \\
  $\gamma$ & \bt {c} 60$^{\rm o} \pm 18^{\rm o}$,\\
 120$^{\rm o} \pm 18^{\rm o}$ \\ \et &
\bt {c} 42$^{\rm o}$ to 78$^{\rm o}$, \\
 102$^{\rm o}$ to 138$^{\rm o}$ \\ \et  &
$42^{\rm o}$ to $ 78^{\rm o}$  & $42^{\rm o}$ to $ 78^{\rm o}$ 
 \\  &&  & & \\
  ${\bar \rho}$ &- 0.334 to 0.334 & - 0.334 to 0.334 &
  0.074 to 0.311  & 0.074 to 0.311 \\&& &   & \\
  ${\bar \eta}$ & 0.235 to 0.439 & 0.235 to 0.439 &
0.235 to 0.439  & 0.244 to 0.439\\
\hline
\end{tabular} 
\caption{The results regarding $ \delta$, $|V_{td}|$, $\alpha$,
 $\beta$, $\gamma$,  ${\bar \rho}$
and ${\bar \eta}$ using the ``future" values:
$|V_{cs}|=  0.996 \pm 0.013 $,
$|V_{ub}/V_{cb}|=0.090 \pm 0.10$ and
$ F_{B_d}\sqrt{B_{B_d}}=0.230 \pm 0.010$GeV.}
\label{tabfuture}
\end{table}


\begin{thebibliography}{99}
\bibitem{paper2} Monika Randhawa, V. Bhatnagar, P. S. Gill and 
M.Gupta, Mod. Phys. Letts {\bf A15}, 2363(2000).   

\bibitem{paper5} Monika Randhawa and Manmohan Gupta,
Phys. Letts. {\bf B516}, 446(2001).

\bibitem{babar}  David G. Hitlin, BABAR Collaboration, hep-ex/0011024;
B. Aubert {\it et al.}, BABAR Collaboration,
Phys. Rev. Lett. {\bf 86}, 2515(2001).

\bibitem{belle}Hiroaki Aihara, BELLE Collaboration, hep-ex/0010008;
 A. Abashian {\it et al.},
 BELLE Collaboration,
Phys. Rev. Lett. {\bf 86}, 2509(2001).

\bibitem{babarlat}  B. Aubert {\it et al.},
 BABAR Collaboration,
Phys. Rev. Lett. 87, 091801(2001).

\bibitem{burasrev}
Andrzej J. Buras, hep-ph/0101336 and references therein.

\bibitem{parodi} M. Ciuchini, G. D'Agostini, E. Franco, 
V. Lubicz, G. Martinelli, F. Parodi, P. Roudeau and 
A. Stocchi, JHEP {\bf 0107}, 013(2001).

\bibitem{hocker} A. Hocker, H. Lacker, S. Laplace and
F. Le Diberder, Eur. Phys. J. {\bf C21}, 225(2001).

\bibitem{kagan} A. L. Kagan and M. Neubert,
Phys. Lett. {\bf B492}, 115(2000).

\bibitem{silva} J. P. Silva and L. Wolfenstein,
Phys. Rev. {\bf D63}, 056001(2001).

\bibitem{nxb} G. Eyal, Y. Nir and G. Perez, JHEP {\bf 0008},
 028(2000);
 Z. Z. Xing, hep-ph/0008018;
 Y. Nir, hep-ph/0008226;
A. J. Buras and R. Buras,
Phys. Lett. {\bf B501}, 223(2001);
 A. Masiero, M. Piai and O. Vives, 
Phys. Rev. {\bf D64}, 055008(2001).

 \bibitem{hurth} T. Hurth, J. Phys. {\bf G27}, 1277(2001);
 T. Hurth and T. mannel, hep-ph/0109041.

\bibitem{gupta}   Manmohan Gupta and P. S. Gill, Pramana {\bf 38},
 477(1992);  P. S. Gill and Manmohan Gupta, Mod. Phys.
 Lett. {\bf A13}, 2445(1998).

\bibitem{rosner} M. Gronau and J. L. Rosner, Phys. Rev. Lett.
 {\bf 76}, 1200(1996);
 J.L. Rosner, Braz. J. Phys. {\bf 31},  147(2001);
 J. Ellis, Nucl. Phys. Proc. Suppl. {\bf 99A}, 331(2001);
H. Fritzsch and Z. Z. Xing, Nucl. Phys. {\bf B556}, 49(1999).

\bibitem{gronau}  R. Fleischer, hep-ph/0011323;
 A. Ali, in Proc. of the 13th Topical Conference 
on Hadron Collider Physics, TIFR, Mumbai, India (1999);
Eur. Phys. J. {\bf C9}, 687(1999); 
I. I. Bigi and A. I. Sanda, hep-ph/9909479.

\bibitem{peccei} R.D. Peccei, hep-ph/9909236; 
hep-ph/0004152;
John Swain and Lucas Taylor, Phys. Rev. D58, 093006(1998);
Stefan Herrlich and Ulrich Nierste, Phys. Rev. {\bf D52}, 6505(1995);
S. Mele, hep-ph/9808411, Proceedings of workshop on CP
 violation, Adelaide, Australia; Phys Rev. {\bf D59}, 113011(1999).

\bibitem{wolf} L. Wolfenstein, Phys. Rev. Lett. {\bf 51}, 1945(1983).

\bibitem{jarlskog} CP violation, Ed. L. wolfenstein, North Holland,
 elsevier Science Publishers B.V., 1989;
 CP violation, Ed. C. Jarlskog, World Scientific Publishing Co. Pte.
  Ltd, 1989.

\bibitem{pdg} D.E. Groom et. al., Particle Data group, Euro. Phys.
 J. {\bf C15}, 1(2000).

\bibitem{branco} G.C. Branco and L. Lavoura, Phys. Lett. {\bf B208},
 123(1988).

\bibitem{bellelat} K. Abe, et al, BELLE Collaboration,
Phys. Rev. Letts.  87, 091802(2001).

\bibitem{buchalla} G. Buchalla, Andrzej J. Buras, M. E. Lautenbacher,
 Rev. Mod. Phys. {\bf 68}, 1125(1996).

\bibitem{lep1} J. Drees, hep-ex/0110077.

 \bibitem{lel} L. Lellouch, Nucl. Phys. Proc. Suppl. {\bf 94},
 142(2001).

\bibitem{hn} S. Herrlich and U. Nierste,
Nucl. Phys. {\bf B419}, 292(1994).

\bibitem{lep} The LEP B Oscillation Working Group, \\
http://lepbosc.web.cern.ch/LEPBOSC/, LEPBOSC 98/3.

\end{thebibliography}
\end{document}